\documentclass[floatfix,aps,twocolumn]{revtex4-1}

\usepackage{amsmath,amssymb,bm,dcolumn,float,Macros,mathrsfs,subfigure}
\usepackage[dvips]{graphicx}
\usepackage{color}
\def\fid{\rm fid}
\def\fnl{f\rom{NL}}
\def\zm{z\rom{m}}
\def\Ng{N\rom{g}}
\def\fsky{f\rom{sky}}
\begin{document}


\title{
Magnification effect on the detection of primordial non-Gaussianity from 
photometric surveys 
}

\author{
Toshiya Namikawa
}
\email{
namikawa@utap.phys.s.u-tokyo.ac.jp
}
\affiliation{
Department of Physics, Graduate School of Science, The University of 
Tokyo, Tokyo 113-0033, Japan
}
\author{
Tomohiro Okamura
}
\email{
t-okamura@astr.tohoku.ac.jp
}
\affiliation{
Astronomical Institute, Tohoku University,Sendai 980-8578, Japan
}
\author{
Atsushi Taruya
}
\email{
ataruya@utap.phys.s.u-tokyo.ac.jp
}
\affiliation{
Research Center for the Early Universe, School of Science, The University 
of Tokyo, Bunkyo-ku, Tokyo 113-0033, Japan
}
\affiliation{
Institute for the Physics and Mathematics of the Universe, The University 
of Tokyo, Kashiwa, Chiba 277-8568, Japan 
}

\date{
February, 2011
}

\begin{abstract}
We present forecast results for constraining the primordial 
non-Gaussianity from photometric surveys through a large-scale 
enhancement of the galaxy clustering amplitude. In photometric surveys,  
the distribution of observed galaxies at high redshifts suffers from the 
gravitational-lensing magnification, which systematically alters the 
number density for magnitude-limited galaxy samples. We estimate size of 
the systematic bias in the best-fit cosmological parameters caused by 
the magnification effect, particularly focusing on the primordial 
non-Gaussianity. For upcoming deep and/or wide photometric surveys like 
HSC, DES and LSST, the best-fit value of the non-Gaussian parameter, 
$\fnl$, obtained from the galaxy count data is highly biased, and the 
true values of $\fnl$ would typically go outside the $3$-$\sigma$ error 
of the biased confidence region, if we ignore the magnification effect 
in the theoretical template of angular power spectrum. The additional 
information from cosmic shear data helps not only to improve the 
constraint, but also to reduce the systematic bias. As a result, the 
size of systematic bias on $\fnl$ would become small enough compared to 
the expected $1$-$\sigma$ error for HSC and DES, but it would be still 
serious for deep surveys with $z_{\rm m}\gtrsim1.5$, like LSST. 
Tomographic technique improves the constraint on $\fnl$ by a factor of 
$2$-$3$ compared to the one without tomography, but the systematic bias 
would increase. 
\end{abstract}

\maketitle

\section{
\label{sec1}
Introduction
}

Any hints on primordial non-Gaussianity would be fruitful to clarify the 
generation mechanism for primordial density fluctuations in the early 
stage of the Universe. Since the single-field slow-roll inflationary 
scenario predicts nearly Gaussian fluctuations (e.g., 
\cite{Allen:1987PhLB,Salopek:1990PhRvD,Falk:1992zg,Gangui:1993tt,
Maldacena:2002vr,Acquaviva:2002ud,Bartolo:2004if}), a detection of large 
primordial non-Gaussianity will rule out the simplest inflationary model 
and provides us a new insight into the physics in the early universe. 

Traditional and popular method to detect primordial non-Gaussianity is 
to measure the three-point correlations of statistical fields (e.g., 
\cite{Komatsu:2001rj,Babich:2004yc,Sefusatti:2007ih,Liguori:2007sj,
Komatsu:2009kd,Jeong:2009vd}). This correlation vanishes in the Gaussian 
fields, and non-vanishing signals of the three-point correlations would 
provide information on primordial non-Gaussianity. 

On the other hand, recent numerical and theoretical studies (e.g., 
\cite{Dalal:2007cu,Desjacques:2008vf,Verde:2009hy,Nishimichi:2009fs,
Taruya:2008pg,Pillepich:2008ka,Giannantonio:2009ak,Grossi:2009an,
Matarrese:2008nc,Schmidt:2010gw,
Liguori:2007sj,Desjacques:2010nn,Desjacques:2010jw}) have revealed that 
the local-type 
non-Gaussianity, originating from the non-linear dynamics of scalar 
fields on super horizon scales, can induce a large-scale enhancement in 
the galaxy clustering amplitude. In the local-type non-Gaussianity, the 
primordial fluctuations characterized by the Bardeen potential, 
$\Phi(\bm{x})$, are described by the Taylor expansion of Gaussian field 
$\phi(\bm{x})$ as (e.g., \cite{Bullock:1998mi,Gangui:1993tt,
Verde:1999ij,Komatsu:2001rj}): 
\beq 
  \Phi(\bm{x}) = \phi(\bm{x}) + \fnl(\phi^2(\bm{x})-\ave{\phi^2}).
\eeq 
The non-vanishing parameter, $\fnl\not=0$, implies a departure from the 
Gaussian statistics, and even a small value of $\fnl$ has been found to 
produce a scale-dependent galaxy bias, which is prominent on large 
scales and at high redshifts. With a help of this property, the 
constraint on primordial non-Gaussianity has been obtained recently, 
combining photometric and spectroscopic surveys 
\cite{Slosar:2008hx,Afshordi:2008ru,Xia:2010pe}, and it turned out that 
the results are rather comparable to that from CMB observation. This 
constraint mainly comes from the quasar data obtained from photometric 
surveys, which are wider and deeper than spectroscopic surveys with a 
limited observation time. In this respect, wide and deep photometric 
surveys planned to start in the near future such as Subaru Hyper 
Suprime-Cam (HSC) survey \cite{HSC}, Dark Energy Survey (DES) 
\cite{DES}, and Large Synaptic Survey Telescope (LSST) survey 
\cite{LSST}) would provide a more stringent constraint on primordial 
non-Gaussianity. 

In those photometric surveys, the observed galaxy distribution at high 
redshifts often suffers from the magnification effect due to the 
gravitational lensing, which apparently changes the number density of 
observed galaxies \cite{Scranton:2005ci,Hildebrandt:2009ez,Wang:2011ur}. 
As increasing redshift, since the amplitude of the density fluctuations 
becomes small and the observed angular separation between any pairs of 
two sources decreases, the galaxy auto-angular power spectrum is shifted 
to small scales with the amplitude decrease. 
On the other hand, the lensing contribution on the 
angular correlations become significant at higher redshifts. 
As a result, the contribution of the magnification effect on the 
galaxy auto correlations is expected to be significant not only at 
high redshifts, but also on large angular scales. 
We thus naively expect that the magnification effect can mimic the 
scale-dependent galaxy bias, and ignoring magnification effect in the 
theoretical template for angular correlations would lead to a biased 
estimation of primordial non-Gaussianity. 

In this paper, we study the potential impacts of the magnification 
effect on constraining primordial non-Gaussianity from the upcoming 
photometric surveys. There are several studies forecasting constraints 
on primordial non-Gaussianity through the scale-dependent galaxy bias 
(e.g., \cite{Carbone:2008iz,Afshordi:2008ru,Carbone:2010sb,
Takeuchi:2010bc}). Here, we pay a particular attention to the 
magnification effect, and quantitatively evaluate the systematic biases 
arising from the incorrect treatment of the magnification effect in 
estimating the non-Gaussianity parameter $\fnl$, which has never been 
considered in the previous forecast studies. Further, we discuss the 
role of the cross correlation statistics between galaxy number density 
and other observables such as cosmic shear. This has been also never 
investigated in previous works, since the cross correlation signals are 
basically insensitive to the primordial non-Gaussianity compared to the 
galaxy auto correlations. However, we found that the cosmic shear-galaxy 
count cross-correlations have large signal-to-noise ratios, and help not 
only to improve the constraint on $\fnl$, but also to reduce the 
systematic bias. 
 
This paper is organized as follows. In Sec.\ref{sec2}, we briefly review 
the scale-dependent galaxy bias induced by the primordial 
non-Gaussianity, and describe how the magnification effect changes the 
number density of observed galaxies. We then give the formalism for the 
angular power spectra obtained from photometric galaxy surveys. In 
Sec.\ref{sec3}, we explicitly compute the angular power spectra and 
calculate the signal-to-noise ratios for auto and cross power spectra 
of galaxy count and cosmic shear. In Sec.\ref{sec4}, based on the Fisher 
matrix formalism, we quantitatively estimate the impact of magnification 
effect on the detection of primordial non-Gaussianity, particularly 
focusing on three representative surveys, i.e., HSC, DES and LSST. 
Finally, Sec.\ref{sec5} is devoted to the summary and discussion. 

Throughout the paper, all the angular power spectra are computed from 
the modified version of cosmological Boltzmann code, CAMB 
\cite{Lewis:1999bs}, with the following set of cosmological parameters 
assuming a flat Lambda-CDM model, which is consistent with WMAP7 results 
\cite{Komatsu:2010fb}; $\Omega\rom{b}h^2=0.022$, $\Omega\rom{m}h^2=0.13$, 
$\Omega_{\Lambda}=0.72$, $n\rom{s}=0.96$, $A\rom{s}=2.4\times10^{-9}$, 
$\tau=0.086$, and $w=-1$, for the density parameters of baryon and 
matter, dark energy density, scalar spectral index, scalar amplitude at 
$k=0.002$ Mpc$^{-1}$, reionization optical depth, dark-energy 
equation-of-state parameter, respectively. Unless otherwise stated, 
non-Gaussian parameter is set to $f\rom{NL}=0$. The non-linear power 
spectrum is computed according to the fitting formula given in 
Ref.\cite{Smith03}. 

\section{
\label{sec2}
Probing primordial non-Gaussianity from photometric surveys 
}


\subsection{
Primordial non-Gaussianity imprinted on galaxy bias
}

In the presence of local-type primordial non-Gaussianity, recent 
numerical and theoretical studies on the clustering of halos/galaxies 
(e.g., \cite{Dalal:2007cu,Desjacques:2008vf,Verde:2009hy,Nishimichi:2009fs,
Taruya:2008pg,Pillepich:2008ka,Giannantonio:2009ak,Grossi:2009an,
Matarrese:2008nc,Schmidt:2010gw}) 
show that there appears a scale-dependent enhancement of the clustering 
amplitude on very large scales. Theoretically, the scale-dependent 
property of the halo/galaxy bias can be explained by a tight correlation 
between long-wavelength and short-wavelength modes, which usually 
vanishes in the Gaussian case. Especially, the modulation of 
short-wavelength modes responsible for forming halos is induced by the 
Newton potential or Bardeen potential $\Phi(\bm{x})$. This fact leads to 
a strong scale-dependence for the fluctuations of halo/galaxy number 
density on large scales, and in Fourier space, we obtain 
\cite{Slosar:2008hx,Reid:2010vc} 
\beq 
  g(\bm{k},z) = [b\rom{G} + \Delta b(k,z)]\delta(\bm{k},z) , \label{gk}
\eeq 
where $g(\bm{k},z)$ and $\delta(\bm{k},z)$ are the fluctuations of galaxy 
number density and matter density fluctuations, respectively, and 
the quantity $b\rom{G}$ implies the galaxy bias in the case of Gaussian 
initial condition. The function $\Delta b(k,z)$ represents the 
non-Gaussian correction, which is given by 
\cite{Slosar:2008hx,Reid:2010vc} 
\beq 
  \Delta b(k,z) = \fnl A\rom{NG}\frac{3\Omega\rom{m}H_0^2}
    {k^2\mcT(k)D(z)} \label{Deltab}. 
\eeq 
Here, the quantity $\Omega\rom{m}$ is the matter energy density, $H_0$ 
denotes the Hubble parameter at present, $D(z)$ is the linear growth 
rate, and $\mcT(k)$ is the transfer function for linear matter density 
fluctuations, which is set to unity in the limit $k\to 0$. Thus, in the 
large-scale limit ($k\to 0$), the second term in Eq.(\ref{gk}) becomes 
dominant, and the enhancement of clustering amplitude is prominent in a 
scale-dependent way. Since this term is inversely proportional to the 
growth rate $D(z)$, non-Gaussian correction becomes also significant at 
higher redshifts. 

Assuming the universality of mass function, the quantity, $A\rom{NG}$, 
gives $\delta\rom{c}(b\rom{G}-1)$ \cite{Dalal:2007cu,Slosar:2008hx}, and 
the quantity $\delta\rom{c}=1.68$ 
is the critical density for a spherical collapse. As advocated by several 
papers, however, the quantity $A\rom{NG}$ would not be simply related to 
the halo mass function, but depends on the merger history of halo/galaxy 
samples \cite{Reid:2010vc}. In other words, we may have to determine 
$A\rom{NG}$ from the observational data in practice. If this is the case, 
the non-Gaussian parameter $\fnl$ would be completely degenerated 
with the quantity $A\rom{NG}$, and we need additional information on 
$\fnl$ like the galaxy bispectrum in order to break the degeneracy. Our 
primary focus here is to explore the impact of magnification effect on 
the non-Gaussian parameter, and we simply assume 
$A\rom{NL}=\delta\rom{c}(b\rom{G}-1)$ in the subsequent analysis. The 
influence of the magnification effect is a generic issue to constrain 
$\fnl$ from the photometric surveys, and we expect that the results in 
the paper are also applicable to the case to combine other observations 
in breaking the degeneracy with $A\rom{NG}$. 

\subsection{
Magnification effect on galaxy number density
}

The number density of galaxies obtained from photometric surveys often
suffers from the magnification effect by the weak gravitational lensing 
of the large-scale structure (e.g., \cite{Moessner:1997qs,
Bartelmann:1999yn,Matsubara:2000pr}). Since the gravitational lensing 
changes the apparent magnitude and the area of the patch in the observed 
sky, it also changes the observed galaxy number density. Denoting the 
zero-mean fluctuations of the observed galaxy number density along a 
direction $\hat{\bm{\theta}}$ at redshift $z$ by 
$n(\hat{\bm{\theta}},z)$, we have \cite{Matsubara:2000pr} 
\beq 
  n(\hat{\bm{\theta}},z) 
    = g(\hat{\bm{\theta}},z) + (5s(z)-2)\kappa(\hat{\bm{\theta}},z) 
    \label{mu} , 
\eeq
where the quantity $\kappa(\hat{\bm{\theta}},z)$ is the lensing 
convergence at the position of source galaxy, and characterizes the 
change of the size of images. The convergence is given by 
\cite{Matsubara:2000pr} 
\beq 
  \kappa(\hat{\bm{\theta}},z) = \frac{3\Omega_mH_0^2}{2}
    \int_0^{\chi(z)}d\chi\frac{\chi(\chi(z)-\chi)}{\chi(z)}
    \delta(\chi\hat{\bm{\theta}},\chi) ,
\eeq 
with the function $\chi(z)$ being the comoving distance. The second 
($5s(z)\kappa$) and third term ($-2\kappa$) in Eq.~(\ref{mu}) arise from 
the modification of apparent magnitude and the area of the patch in the 
observed sky by lensing, respectively. 

The magnitude of the lensing effect depends on the slope parameter, 
$s(z)$. Denoting the number of galaxies at redshift $z$, brighter than 
the magnitude $m$ by $N(z,<m)$, the quantity $s(z)$ is defined by 
\cite{Matsubara:2000pr}
\beq 
  s(z) \equiv \frac{d\log_{10} N(z,<m)}{dm} . 
\eeq

Note that in addition to the correction in Eq.~(\ref{mu}), there exists 
another possible contribution related to the lensing effect, which has 
been addressed in Refs.~\cite{Schmidt:2009rh,Schmidt:2009ri}. That is, 
the observed galaxies are selected according not only to the magnitude 
cut, but also to the size cut, and the latter also alters the galaxy 
number density. Nevertheless, the effect of size cut is basically 
proportional to the lensing convergence, and can be incorporated into 
the expression (\ref{mu}), with a slight change of the meaning of slope 
index, $s(z)$. In this respect, the results in the present paper is 
general, and applicable to the case taking account of the size cut.

\subsection{
The angular power spectra
}

The angular power spectra are the fundamental statistical quantity 
obtained from the photometric survey, and have a rich cosmological 
information. Here we write down the expressions for angular power 
spectra of galaxy number counts and cosmic shear obtained from 
photometric surveys. 

The galaxy number density observed via photometric survey is projected 
onto the two-dimensional sky, and redshift information is obtained by 
dividing photometric galaxy samples into several subsamples binned with 
redshifts. With the redshift distribution of galaxies in $i$-th bin, 
$N_i(z)$, the two-dimensional distribution of observed galaxies in 
$i$-th bin, $n_i(\hat{\bm{\theta}})$, are given by 
\beq 
  n_i(\hat{\bm{\theta}}) 
    = \int dz \frac{N_i(z)}{\overline{N}_i}n(\hat{\bm{\theta}},z) .
\eeq
The quantity $\overline{N}_i$ is the average number density per square 
arcminute in $i$-th bin, defined by 
\beq 
  \overline{N}_i = \int_0^{\infty} dz_s N_i(z_s) .
\eeq 
On the other hand, the cosmic shear is measured from the ellipticity of 
each galaxy image. Using the photometric redshift information, we can 
also divide the estimated shear into several redshift bins. We denote 
the cosmic shear field in the $i$-th redshift bin by 
$\gamma_i(\hat{\bm{\theta}})$. Then, the angular power spectra between 
the observables, $X$ and $Y$ ($X$ and $Y$ are either of $\gamma_i$ or 
$n_j$) are given by the following expression: 
\beq
  C_{\ell}^{XY}=\frac{2}{\pi}\int \frac{dk}{k^2}\, P\rom{init}(k)
  \Delta_{\ell}^X(k)\Delta_{\ell}^Y(k) , 
  \label{PS}
\eeq
where $P\rom{init}(k)$ is the matter power spectrum at an early time and 
$k$ is the Fourier wave number. The functions $\Delta^X_{\ell}(k)$ and 
$\Delta^Y_{\ell}(k)$ are one of the following 
\cite{Hu:2000ee,LoVerde:2006cj,Kostelecky:2008iz}:
\al{
  \Delta^{\gamma_i}_{\ell}(k) 
    &= \sqrt{\frac{(\ell+2)!}{(\ell-2)!}}\,
      \,\mcP_{\ell}(k;N_i(z)), 
    \label{Delta-S} \\
  \Delta^{n_i}_{\ell}(k) 
    &= k^2 \int dz [b_G+\Delta b(k,z)]\frac{N_i(z)}{\overline{N}_i}
      D(z)j_{\ell}(k\chi(z)) 
      \notag \\
      &\quad + (5 s_i-2)\ell(\ell+1)\mcP_{\ell}(k;N_i(z)). 
    \label{Delta-n}
}
The function $j_{\ell}$ is the spherical Bessel function and 
$\mcP_{\ell}(k;N_i(z))$ is defined by 
\al{ 
  \mcP_{\ell}(k;N_i(z)) &= \frac{3\Omega_{\rm m} H_0^2}{2}
    \int_0^{\infty}dz_s\,\frac{N_i(z_s)}{\overline{N}_i} \notag \\ 
    &\times \int_0^{\chi(z_s)}\!\!\! d\chi\, 
    \frac{\chi(z_s)-\chi}{\chi(z_s)\chi}
    \,\frac{D(z(\chi))}{a(\chi)}\,j_{\ell}(k\chi)
    \label{P} .
} 
The quantity $s_i$ is the slope index in the $i$-th redshift bin. 
Although the slope index seems to have a strong redshift 
dependence \cite{LoVerde:2006cj,Hui:2007cu}, we here assume the constant 
slope index within each redshift bin, and study the effect of time 
varying slope index. 

Note that for the photometric redshift determination, the uncertainty 
arising from the photometric redshift error is crucial for the 
cosmological analysis \cite{Amara:2006kp}. To mimic this effect, we 
suppose that the photometric redshift estimates are distributed as a 
Gaussian with rms fluctuation $\sigma(z)$. Then the actual redshift 
distribution for $i$-th galaxy subsamples over the range, 
$z_{i-1}<z<z_i$, is related to the redshift distribution of galaxies, 
$N(z)$, as \cite{Hu:2004yd} 
\beq 
  N_i(z)=\frac{1}{2}\,N(z)\left[{\rm erfc}
  \left(\frac{z_{i-1}-z}{\sqrt{2}\sigma(z)}\right)-{\rm erfc}
  \left(\frac{z_i-z}{\sqrt{2}\sigma(z)}\right)\right] 
  \label{n-z}, 
\eeq
where the function ${\rm erfc}(x)$ is the complementary error function 
defined by
\beq 
  {\rm erfc}(x)\equiv \frac{2}{\sqrt{\pi}}\int_x^{\infty}dz\exp(-z^2). 
\eeq

\section{
\label{sec3}
Magnification effect on angular power spectra and signal-to-noise ratio
}

In this section, adopting a simple model of time evolution for bias 
parameter $b\rom{G}$, we compute the angular power spectra, and show how 
the magnification effect changes the amplitude of angular power spectra.  
Also, based on the fiducial setup of future photometric surveys, we 
estimate the signal-to-noise ratios for auto- and cross-power spectra of  
galaxy counts and cosmic shear. 

In Eq.~(\ref{gk}), while the parameter $b\rom{G}$ is assumed to be 
scale-independent on large-scales, it would manifest a strong time 
dependence. We characterize this by introducing the following function 
(e.g., \cite{Fry:1996ApJ,Tegmark:1998wm,Yamamoto:2006yv}): 
\beq
  b\rom{G} = b_0 + \frac{b_z}{D(z)} \label{bias}. 
\eeq
Here, we set $b_0=1.5$ and $b_z=2.0$ for fiducial values of the galaxy 
bias parameters. Since the photometric redshift information is available 
for most of the photometric surveys, we employ the tomographic technique,  
and divide all the galaxy samples into the three redshift subsamples.  
The redshift ranges for each bin are chosen as, $z<0.739$, 
$0.739<z<1.14$, and $1.14<z$, so that each redshift bin has the same 
number of galaxies. We assume the redshift distribution of galaxies 
$N(z)$ as (e.g., \cite{Yamamoto:2007gd}) 
\beq 
  N(z) = \Ng\frac{3z^2}{2(0.64 \zm)^3} 
    \exp\left[-\left(\frac{z}{0.64 \zm}\right)^{3/2}\right] , \label{Nz}
\eeq
where the quantities $\Ng$ and $\zm$ are the total number of galaxies 
per square arcminute and mean redshift. As typical values of the 
upcoming deep surveys, we set $\Ng$ to $35$ arcmin$^{-2}$ and $\zm=1.0$. 
As for the photo-$z$ error, we adopt the simple scaling relation 
\cite{Hu:2004yd}:
\beq 
  \sigma(z) = 0.03\,(1+z). \label{sigmaz} 
\eeq
Finally, the influence of magnification effect on the angular power 
spectra depends on the slope parameter, for which we set $s_1=0.5$, 
$s_2=1.0$, and $s_3=1.5$, close to the recently estimated values from 
the observations \cite{LoVerde:2006cj,Hui:2007cu}. 



\subsection{
Magnification effect on angular power spectra
\label{sec3.1}
}

\begin{figure}[t]
\bc
\includegraphics[width=85mm,clip]{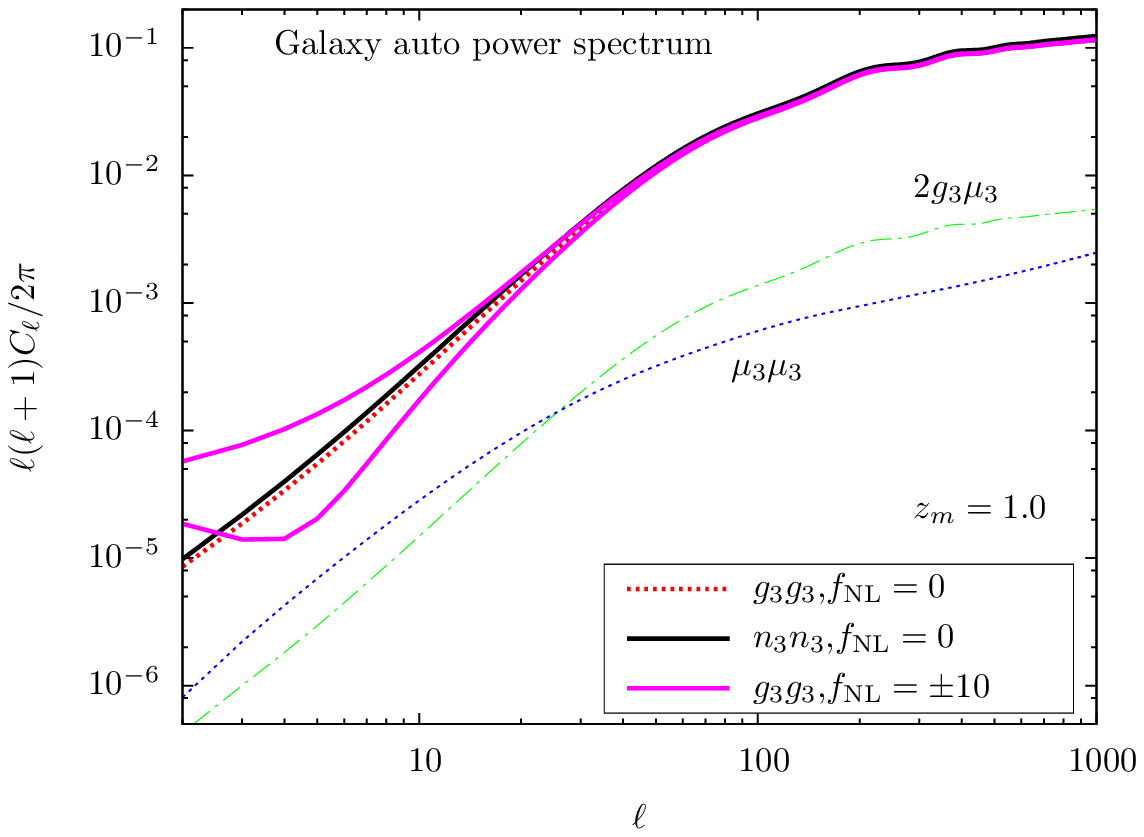}
\includegraphics[width=85mm,clip]{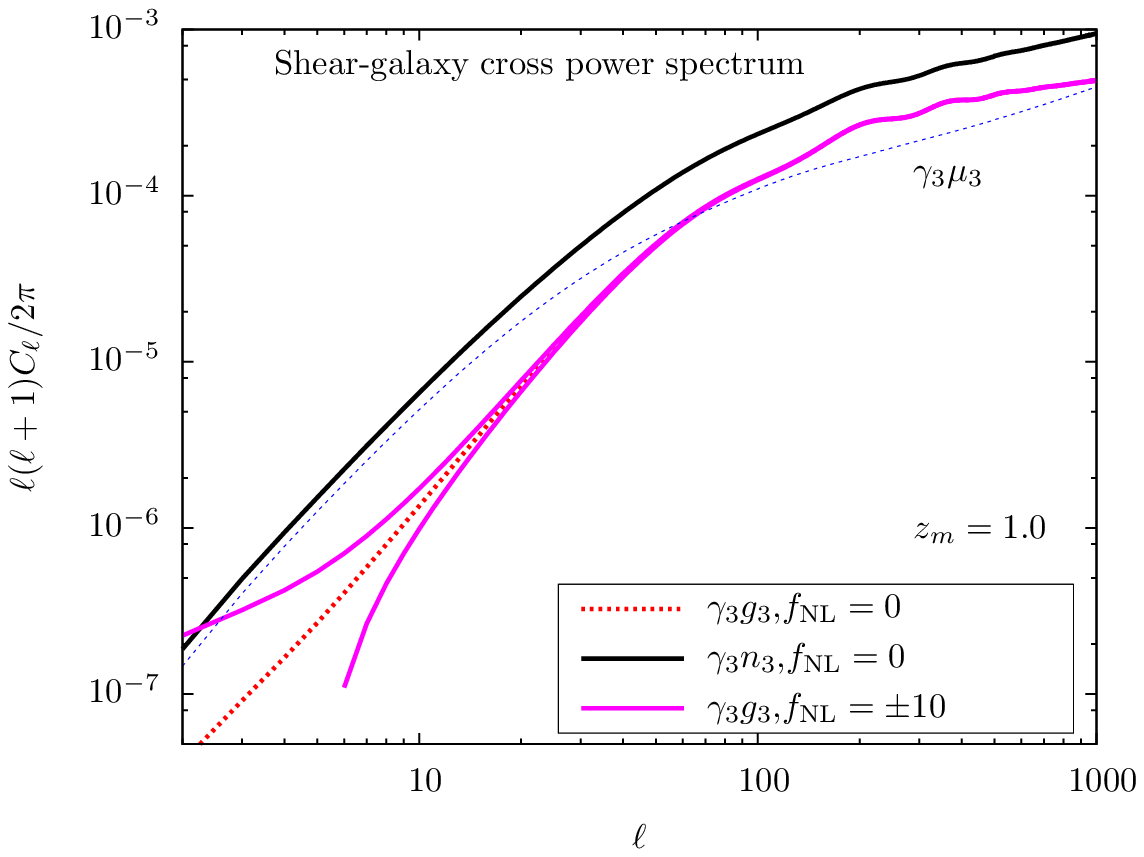}
\ec 
\caption{\label{Cl}
The galaxy auto-power spectrum $C_{\ell}^{nn}$ (top) and the 
shear-galaxy cross-power spectrum $C_{\ell}^{\gamma n}$ (bottom).
We plot the angular power spectra in the presence/absence of the 
magnification effect (black solid/red dotted) for the Gaussian initial 
condition, $\fnl=0$, and also show the angular power spectra 
in the absence of the magnification effect in the non-Gaussian case, 
$\fnl=\pm10$ (magenta solid). 
For comparison, we plot the contribution of the magnification 
effect ($2g_3\mu_3$, and 
$\mu_3\mu_3$ in the top and $\gamma_3\mu_3$ in the bottom panel). 
}
\end{figure}

\begin{figure}[t]
\bc
\includegraphics[width=85mm,clip]{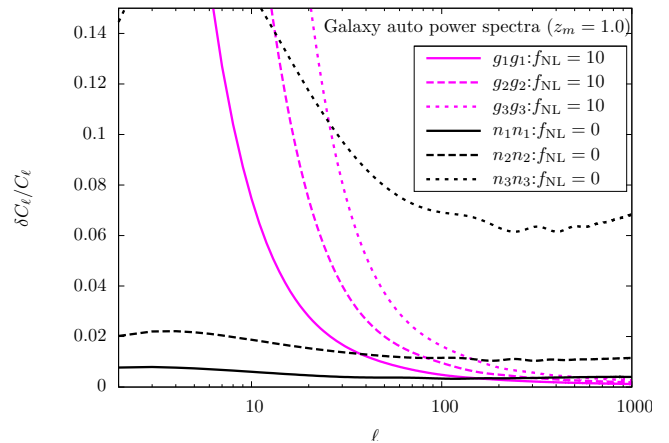}
\ec 
\caption{\label{Clz}
The dependence of galaxy auto-power spectra on the parameter $\fnl$
and the magnification effect for each tomographic bin. We define the 
fractional difference of galaxy auto-power spectra, 
$\delta C_{\ell}/C_{\ell}$ which is given in two cases: 
$C_{\ell}^{g_ig_i}(\fnl=10)/C_{\ell}^{g_ig_i}(\fnl=0)-1$ 
(magenta lines) and 
$C_{\ell}^{n_in_i}(\fnl=0)/C_{\ell}^{g_ig_i}(\fnl=0)-1$ 
(black lines) . 
The solid, dashed and dotted lines represent $(i,j)=(1,1)$, $(2,2)$ and 
$(3,3)$ bin, respectively.
}
\end{figure}

In the presence of the magnification effect, the galaxy auto-power 
spectra, $C_{\ell}^{n_in_j}$, and the shear-galaxy cross-power spectra, 
$C_{\ell}^{\gamma n_i}$, can be separately decomposed into several 
pieces: 
\al{
  C_{\ell}^{n_in_j} 
    &= C_{\ell}^{g_ig_j}+C_{\ell}^{g_i\mu_j} 
    + C_{\ell}^{\mu_i g_j}+C^{\mu_i\mu_j}_{\ell}, \label{gg-decomp}
    \\
  C_{\ell}^{\gamma_i n_j} 
    &= C_{\ell}^{\gamma_ig_j}+C_{\ell}^{\gamma_i \mu_j},
}
where the subscripts $g_i$ and $\mu_i$ respectively represent the 
contribution of the {\it pure} galaxy clustering and magnification, 
which are identified with the first and second terms in 
Eq.~(\ref{Delta-n}). That is, the power spectra involving $g_i$ and 
$\mu_i$ can be computed by simply neglecting the second and first terms 
in Eq.~(\ref{Delta-n}), respectively. 

In Fig.~\ref{Cl}, the galaxy auto and shear-galaxy cross angular power 
spectra ($C_{\ell}^{n_3 n_3}$ and $C_{\ell}^{\gamma_3 n_3}$) are shown. 
We plot the power spectra in the presence/absence of the magnification 
effect for the Gaussian initial condition (black solid/red dotted), 
and also show the power spectra in the absence of the magnification 
effect for the non-Gaussian case, $\fnl=\pm10$ (magenta solid). 
Note that the power spectra plotted here are independent of $N_g$ in 
Eq.~(\ref{Nz}) (see Eqs.~(\ref{Delta-S})-(\ref{P})). 

For the galaxy auto power spectra $C_{\ell}^{n_3 n_3}$, there appears 
two contributions of the magnification effect in Eq.~(\ref{gg-decomp}), 
i.e., $C^{\mu_3\mu_3}_{\ell}$ and $C^{g_3\mu_3}_{\ell}$. As shown in the 
top panel of Fig.~\ref{Cl}, the cross power spectrum between galaxy 
counts and the magnification, $C_{\ell}^{g_3\mu_3}$ (green dot-dashed), 
has similar angular dependence to $C_{\ell}^{g_3g_3}$ (red dotted), and  
it slightly changes the overall amplitude of power spectra. The 
contribution of magnification auto-power spectra, $C_{\ell}^{\mu_3\mu_3}$ 
(blue dotted), has also similar $\ell$-dependence, but the amplitude 
exceeds $C_{\ell}^{g_3\mu_3}$ on large scales. This feature comes from 
the fact that the weak lensing effect is mainly attributed to the growth 
of structure at lower redshifts. In the same manner, in the bottom panel 
of Fig.~\ref{Cl}, the amplitude of shear-galaxy cross-power spectra, 
$C_{\ell}^{\gamma_3g_3}$, is enhanced at low-$\ell$ by the magnification 
effect. Hence, the magnification effect leads to a scale-dependent 
enhancement in the amplitude of angular power spectra, which can mimic 
the effect of primordial non-Gaussianity through the scale-dependent 
galaxy bias. On the other hand, the bias parameters $b_0$ and $b_z$ are 
independent of the scales, and their influences appear not only on large 
scales but also on small scales. Thus, even if treating these as free 
parameters and marginalizing over them, the constraint on $\fnl$ 
seems to be unaffected by the galaxy bias parameters $b_0$ and $b_z$.  

In Fig.~\ref{Clz}, to elucidate the scale-dependent enhancement of the 
power spectra on large scales, we plot the fractional difference of the 
galaxy auto power spectra, $\delta C_{\ell}/C_{\ell}$. Here, we examine 
the two cases: $C_{\ell}^{g_ig_i}(\fnl=10)/C_{\ell}^{g_ig_i}(\fnl=0)-1$ 
(magenta lines) and 
$C_{\ell}^{n_in_i}(\fnl=0)/C_{\ell}^{g_ig_i}(\fnl=0)-1$ (black lines). 
As we expected, the impact of magnification effect is significant at 
higher redshift bins. This is because, as increasing the source 
redshifts, the gravitational lensing becomes significant and the 
amplitude of fluctuations $g$ conversely decreases. The contribution of 
primordial non-Gaussianity is also significant at higher redshifts, 
because the non-Gaussian correction in the scale-dependent galaxy bias 
is proportional to the inverse of growth function, $D(z)$ 
[see Eq.(\ref{Deltab})]. Note, however, that the redshift dependence of 
the magnification effect (magenta lines) is rather different from that 
of the scale-dependent galaxy bias (black lines). This implies that the 
tomographic technique is useful to break the degeneracy between the 
effects of magnification and primordial non-Gaussianity.

\subsection{
Signal-to-noise ratio
\label{sec3.2}
}

\begin{figure}[t]
\bc
\includegraphics[width=85mm,clip]{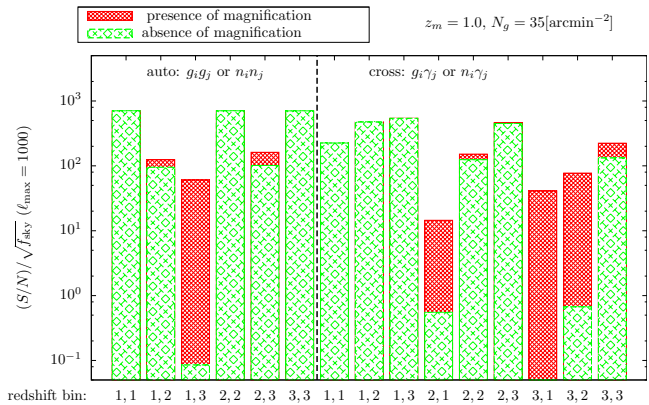}
\ec 
\caption{
\label{SN}
The signal to noise ratio of each power spectra in the presence/absence 
of the magnification effect (red/green) for a survey with the mean 
redshift $\zm=1.0$ and the number density of galaxy $\Ng=35$ 
arcmin$^{-2}$. We show the signal-to-noise ratio for all power spectra 
normalized by $\sqrt{f\rom{sky}}$ since the signal-to-noise ratio is 
proportional to $\sqrt{f\rom{sky}}$. 
}
\end{figure}

Since the magnification effect enhances the amplitude of power spectra 
especially on large scales, the signal-to-noise ratio for angular power 
spectra would be changed. To estimate the size of this, in Fig.~\ref{SN}, 
we plot the signal-to-noise ratios for galaxy auto and shear-galaxy 
cross spectra in the presence or absence of the magnification effect. 
The signal-to-noise ratio, $\mathrm{S/N}$, is defined by 
\al{
  \frac{S}{N}\equiv\sqrt{\sum_{\ell=2}^{\ell\rom{max}}
    \left(\frac{C_{\ell}^{XY}}{\Delta C_{\ell}^{XY}}\right)^2}, 
}
where the quantity $\Delta C_{\ell}^{XY}$ is the statistical error for 
each power spectra, given by 
\al{
  \left(\Delta C_{\ell}^{XY}\right)^{2} 
    = &\frac{1}{(2\ell + 1)\fsky}\left[
    \left(C_{\ell}^{XY}+N_{\ell}^{XY}\right)^{2}
      \right. \notag \\
    &\left.+\left(C_{\ell}^{XX}+N_{\ell}^{XX})
    (C_{\ell}^{YY}+N_{\ell}^{YY}\right) \right] . 
}
Here, the parameter $\fsky$ is the sky coverage of photometric survey,  
and $N_{\ell}^{XY}$ are the noise power spectra, which will be 
later given in Sec.IV [see Eqs.~(\ref{eq:noise_n})-(\ref{eq:noise_gamma})]. 
In Fig.~\ref{SN}, the signal-to-noise ratios are computed with 
$\ell\rom{max}=1000$, and are normalized by $\sqrt{\fsky}$. 

For the cross power spectra which have primarily no statistical 
correlation [i.e., $C^{n_in_j}_{\ell}$ ($i\not=j$) and 
$C^{\gamma_in_j}_{\ell}$ ($i<j$)] \footnote{Due to the photo-$z$ error, 
however, there appears statistical correlation even in the absence of 
magnification effect, although the signal is very weak.}, the 
signal-to-noise ratios are significantly improved by the magnification 
effect. For example, the signal-to-noise ratio of 
$C_{\ell}^{\gamma_2n_3}$ increases by a factor of $\sim 115$. This is 
because the magnification effect leads to a non-vanishing correlation 
between foreground galaxies and the background sources. On the other 
hand, the improvement of the signal-to-noise ratio is relatively small 
for the power spectra which have strong statistical correlation even in 
the absence of the magnification effect 
[i.e., $C^{n_in_j}_{\ell}$ ($i=j$), and 
$C^{\gamma_in_j}_{\ell}$ ($i\geq j$)]. Since the signals of these spectra 
are primarily very large enough to detect, the results indicate that the 
size of the constraint on $f\rom{NL}$ would not be drastically changed 
even if the magnification effect is properly taken into account in the 
data analysis. We finally note that, even in the absence of the 
magnification effect, the signal-to-noise ratios of the shear-galaxy 
cross correlations [$C^{\gamma_in_j}_{\ell}$ ($i\geq j$)] are comparable 
to that of the galaxy auto correlations [$C^{n_in_j}_{\ell}$ ($i=j$)]. 
This implies that, the parameter $\fnl$ would be constrained not only 
from the galaxy auto correlations, but also from the shear-galaxy 
cross correlations.

\section{
\label{sec4}
Magnification effect on the detection of primordial non-Gaussianity
}

In this section, based on the Fisher matrix formalism, we now present the 
forecast constraint on $\fnl$ from photometric surveys, and estimate the 
size of systematic bias from the incorrect treatment of magnification 
effect in the theoretical template of angular power spectra. As 
representative upcoming experiments for wide and/or deep surveys, we 
consider HSC experiment for wide but narrow, DES for shallow but wide, 
and LSST for idealistically deep and wide surveys. 

\subsection{
\label{sec4.1} 
Fisher matrix formalism 
}

Here, we summarize the Fisher matrix formalism used in the subsequent 
analysis, and describe the canonical survey setup for photometric galaxy 
surveys and CMB experiments. 

Given the angular power spectra theoretically parametrized by a set of 
parameters $\bm{p}$, the Fisher matrix for the cosmological parameters 
is written as (e.g., \cite{Tegmark:1996bz})
\al{ 
  F_{ij} 
    &= \sum_{\ell=2}^{\ell\rom{max}}\frac{2\ell+1}{2}\fsky \notag \\ 
    &\times \Tr\left(\bm{C}^{-1}_{\ell}(\bm{p})
    \PD{\bm{C}_{\ell}}{p_i}(\bm{p})\bm{C}^{-1}_{\ell}(\bm{p})
    \PD{\bm{C}_{\ell}}{p_j}(\bm{p})\right)\bigg|_{\bm{p}=\bm{p}^{\fid}}, 
  \label{FisMat} 
}
where the quantity $\bm{C}_{\ell}$ represents the covariance matrix for 
the angular power spectra, $p_i$ is a cosmological parameter which we 
want to estimate, and $\bm{p}^{\fid}$ is the set of fiducial cosmological 
parameters. 

In what follows, using a tomographic technique with the number of 
redshift bin, $N\rom{bin}$, we consider the number density fluctuations 
$n_1,\dots,n_{N\rom{bin}}$ (or $g_1,\dots g_{N\rom{bin}}$ when ignoring 
the magnification effect), and shear fields, 
$\gamma_1,\dots,\gamma_{N\rom{bin}}$, as observables obtained from the 
photometric surveys. Since these observables are rather sensitive to the 
late-time cosmic expansion and/or growth of structure, photometric 
surveys alone cannot give a tight constraint on all the cosmological 
parameters. To break the degeneracy between cosmological parameters and 
improve the constraints, we include the information obtained from the 
primary CMB anisotropies by Planck \cite{Planck}. To be specific, we use 
the temperature ($\Theta$) and (E-mode) polarization ($E$) data for 
primary CMB anisotropies. Denoting the noise power spectra by 
$N_{\ell}^{XY}$, the full covariance matrix, $\bm{C}_{\ell}$, is written 
as 
\beq 
  [\bm{C}_{\ell}]_{ij} = C^{X_iX_j}_{\ell}+N_{\ell}^{X_iX_j}\delta_{ij}, 
  \label{Cov-all} 
\eeq
where $X_i$ and $X_j$ stand for $n_1,\dots,n_{N\rom{bin}}$ 
(or $g_1,\dots,g_{N\rom{bin}}$ in the absence of the magnification 
effect), $\gamma_1,\dots,\gamma_{N\rom{bin}}$, $\Theta$, or $E$. 
The amplitude and shape of the noise spectra $N_\ell^{XY}$ depends on 
the survey design which will be discussed below. 

In the Fisher-matrix analysis, the forecast constraints depend on the 
properties of a photometric survey, characterized by the sky coverage, 
$\fsky$, the mean redshift, $\zm$, and the total number of galaxies per 
square arcminute, $\Ng$. To show how the constraint on $\fnl$ depends on 
the survey design, we compute the Fisher matrix in the three 
representative photometric surveys; HSC for a deep survey ($\fsky=0.05$, 
$\zm=1.0$ and $\Ng=35$ arcmin$^{-2}$) and DES ($\fsky=0.125$, $\zm=0.5$ 
and $\Ng=12$ arcmin$^{-2}$) for a wide imaging surveys, and LSST 
($\fsky=0.5$, $\zm=1.5$ and $\Ng=100$ arcmin$^{-2}$) as an idealistic 
survey, which is deeper and wider enough than the HSC and DES surveys. 
In Table \ref{Instru-G}, we summarize the basic parameters of the survey 
design for three surveys used in the subsequent analysis. 

Let us now consider the noise spectra for each data set. In a 
photometric survey, the main noise source for galaxy counts is the shot 
noise given by 
\beq 
  N_{\ell}^{n_i n_j}=\delta_{ij}\frac{1}{\hat{N}_i} , 
  \label{eq:noise_n}
\eeq
with the quantity $\hat{N}_i$ being the number density of galaxies per 
steradians in $i$-th redshift bin; 
\beq 
  \hat{N}_i = 3600 \,\bar{N}_i\,\left(\frac{180}{\pi}\right)^2 \quad 
    {\rm str}^{-1} . 
\eeq
On the other hand, the noise source for cosmic shear measurement mainly 
comes from the intrinsic ellipticity of galaxies, which is described by
\beq 
  N_{\ell}^{\gamma_i\gamma_j}
    = \delta_{ij}\frac{\ave{\gamma\rom{int}^2}}{\hat{N}_i} . 
\label{eq:noise_gamma}
\eeq
The quantity $\ave{\gamma\rom{int}^2}^{1/2}$ is the rms intrinsic 
ellipticity. We adopt the empirically derived value, 
$\ave{\gamma\rom{int}^2}^{1/2}=0.3$ \cite{Bernstein:2001nz}. 
In all surveys, the galaxy samples are divided into three redshift bins, 
and the ranges of redshift are chosen such that each redshift bin has 
same number of galaxies, $\Ng/3$. The resultant redshift ranges are 
summarized in Table \ref{zm} for each mean redshift $\zm$. 
Note that the noise power spectra of CMB anisotropies, 
$N_{\ell}^{\Theta\Theta}$ and $N_{\ell}^{EE}$, are computed according to 
Eq.~(3.3) of Ref.\cite{Namikawa:2010re}, with the experimental 
specification for Planck \cite{Planck}. 

\begin{table}[t]
\caption{
\label{Instru-G}
Survey design for HSC, DES and LSST, i.e., the sky coverage, $\fsky$, 
the mean redshift, $\zm$, and the number of galaxies per square 
arcminute, $\Ng$. In all surveys, the galaxy samples are divided into 
three redshift bin, and the ranges of redshift are chosen such that the 
each redshift bin has same number of galaxies, $\Ng/3$. The resultant 
redshift ranges are summarized in Table \ref{zm} for each mean redshift 
$\zm$. 
}
\begin{tabular}{ccccccc} \hline 
Survey & $\fsky$ & $\zm$ & $\Ng$ [arcmin$^{-2}$] \\ \hline 
HSC \cite{HSC} & 0.05 ($2000$deg$^2$)  & 1.0 & 35 \\ \hline 
DES \cite{DES} & 0.125 ($5000$deg$^2$) & 0.5 & 12 \\ \hline 
LSST \cite{LSST} & 0.5 ($20000$deg$^2$) & 1.5 & 100 \\ \hline 
\end{tabular}
\end{table}

\begin{table}[t]
\caption{
\label{zm}
The relation between mean redshift, $\zm$, and the redshift ranges of 
$i$-th bin computed in the case with $N\rom{bin}=3$. Using 
Eq.(\ref{n-z}), the redshift ranges are determined such that each 
redshift bin has same number of galaxies.
}
\begin{tabular}{c|ccc} \hline 
$\zm$ & \multicolumn{3}{c}{redshift ranges} \\ \hline 
0.5 & $z<0.369$ & $0.369<z<0.569$ & $0.569<z$ \\ \hline 
1.0 & $z<0.739$ & $0.739<z<1.14$ & $1.14<z$ \\ \hline 
1.5 & $z<1.11$ & $1.11<z<1.72$ & $1.72<z$ \\ \hline 
2.0 & $z<1.48$ & $1.48<z<2.29$ & $2.29<z$ \\ \hline 
\end{tabular}
\end{table}


From the Fisher matrix, the 1-$\sigma$ ($68$ \%C.L.) constraint on a 
cosmological parameter, $\sigma(p_i)$, marginalized over other 
parameters, is given by $\{F_{ii}^{-1}\}^{1/2}$. Number of free 
parameters in the subsequent Fisher analysis is, in total, $13$, i.e., 
$\Omega\rom{b}h^2$, $\Omega\rom{m}h^2$, $\Omega_{\Lambda}$, $n\rom{s}$, 
$A\rom{s}$, $\tau$, $w$, $\fnl$, in addition to the galaxy bias 
parameters ($b_0$ and $b_z$) and slope indices ($s_1,$,$s_2$,$s_3$). 
Note that the number of free parameters is changed to $10$ if we 
incorrectly neglect the magnification effect in the theoretical template 
of angular power spectra. The fiducial values of the parameters are the 
same as those in Sec.\ref{sec3}, and we assume that the photo-$z$ error 
and redshift distribution of galaxies are given in Eq.(\ref{sigmaz}) and 
Eq.(\ref{Nz}), respectively. 

Using the Fisher matrix formalism, we also evaluate the systematic bias 
in the best-fit value of cosmological parameters, $\Delta p_i$, arising 
from the incorrect treatment of the magnification effect in the 
theoretical template. Assuming the Gaussian likelihood function, the 
bias of the best-fit value, $\Delta p_i$, can be estimated from 
\cite{Joachimi:2009fr} 
\al{ 
  \Delta p_i 
    &= \frac{1}{2}\sum_{i,j}\tilde{F}_{ij}\frac{2\ell+1}{2}\fsky 
    \notag \\ 
    &\times \Tr\left(\tilde{\bm{C}}^{-1}(\bm{p})
    \PD{\tilde{\bm{C}}}{p_i}(\bm{p})\widetilde{\bm{C}}^{-1}(\bm{p})
    (\bm{C}(\bm{p})-\tilde{\bm{C}}(\bm{p}))\right)
    \bigg|_{\bm{p}=\bm{p}^{\fid}} , \label{bias-F} 
}
where the covariance matrices $\bm{C}$ and $\tilde{\bm{C}}$ are computed 
with and without the magnification effect in the power spectra, 
respectively. The Fisher matrix $\tilde{F}_{ij}$ is computed by ignoring 
the magnification in angular power spectra. 

\subsection{
\label{sec4.2} 
Results 
}

\begin{figure}[t]
\bc
\includegraphics[width=80mm,clip]{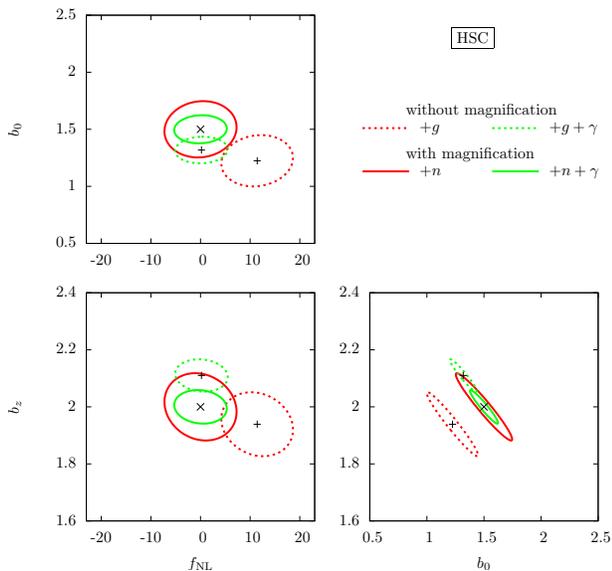} 
\ec 
\caption{
\label{HSC}
The 1-$\sigma$ error contours on $b_0$-$\fnl$ (top left), $b_z$-$\fnl$ 
(bottom left) and $b_z$-$b_0$ (bottom right) planes for HSC. In each 
panel, we show the constraints in the case with (solid) and without 
(dashed) the magnification effect in theoretical template. In each case, 
we show the results obtained from galaxy number counts alone, and 
further including cosmic shear of galaxies. Note that CMB prior 
information from Planck is included in all cases. Information from the 
primary CMB is summed up to $\ell=3000$, and other signals are included 
up to $\ell\rom{max}=1000$. We also note that the fiducial values of 
the galaxy bias parameters and slope indices are chosen as $b_0=1.5$, 
$b_z=2.0$, $s_1=0.5$, $s_2=1.0$ and $s_3=1.5$. 
}
\end{figure}

\begin{figure}[t]
\bc
\includegraphics[width=80mm,clip]{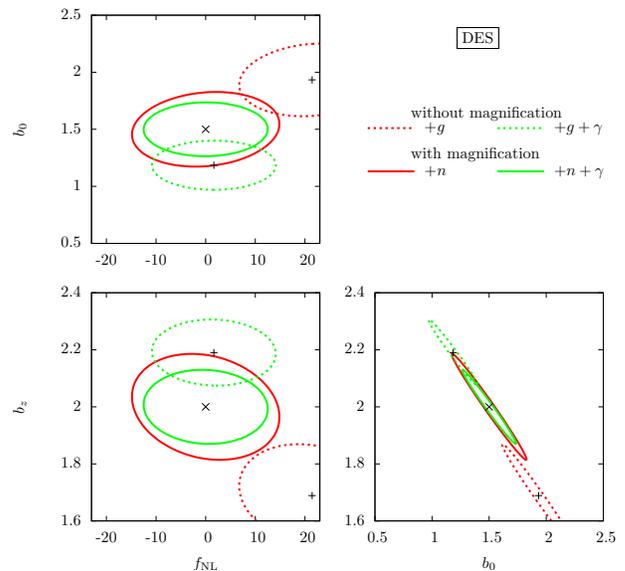} 
\ec 
\caption{
\label{DES}
Same as Fig.\ref{HSC} but for DES. 
}
\end{figure}

\begin{figure}[t]
\bc
\includegraphics[width=80mm,clip]{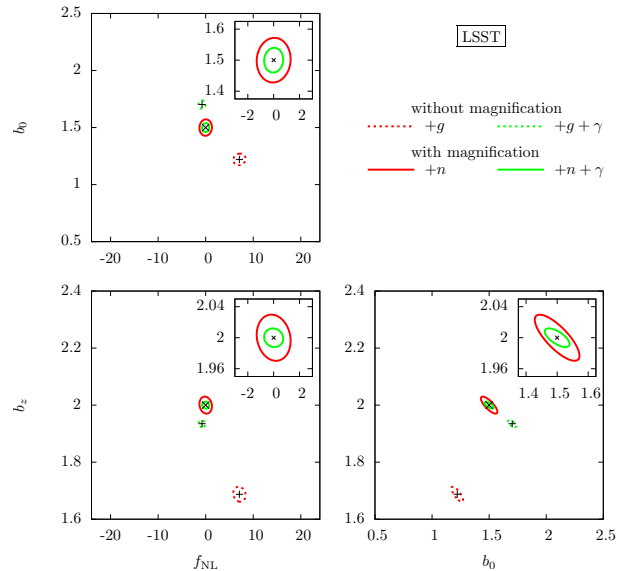} 
\ec 
\caption{
\label{LSST}
Same as Fig.\ref{HSC} but for LSST. 
The inset shows the zoomed error contours around the fiducial values. 
}
\end{figure}

\begin{table}[t]
\caption{
\label{const}
The 1-$\sigma$ constraints, $\sigma(\fnl)$, and the systematic bias, 
$\Delta \fnl$, on the primordial non-Gaussianity, in the case 
with/without the magnification effect in theoretical template. In each 
case, we show the resultant constraints obtained from galaxy number 
counts alone, and further including cosmic shear of galaxies. CMB prior 
information from Planck is included in all cases. We set the maximum 
multipole as $\ell\rom{max}=1000$, but the primary CMB power spectra is 
used up to $\ell=3000$ in the Fisher matrix. 
}
\bc
\begin{tabular}{c|c|c|c|c|c|c} \hline 
 & \multicolumn{2}{c|}{HSC} & \multicolumn{2}{c|}{DES} 
 & \multicolumn{2}{c}{LSST} \\ \hline 
 & $\sigma(\fnl)$ & $\Delta \fnl$ & $\sigma(\fnl)$ & $\Delta \fnl$
 & $\sigma(\fnl)$ & $\Delta \fnl$ \\ \hline 
$+n$  & 4.8 & - & 9.8 & - & 0.86 & - \\ 
$+g$  & 4.7 & 11 & 9.7 & 21 & 0.86 & 7.1 \\ \hline
$+n+\gamma$ & 3.5 & - & 8.2 & - & 0.49 & - \\ 
$+g+\gamma$ & 3.5 & 0.19 & 8.2 & 1.7 & 0.49 & -0.76 \\ 
\hline
\end{tabular}
\ec
\end{table}

\subsubsection{
Estimation of primordial non-Gaussianity for representative surveys 
\label{Sec.4.2.1}
}

Here we present forecast results for the constraint on primordial 
non-Gaussianity, especially focusing on the following cases: 
\bi 
  \item $+n$: using galaxy counts alone {\it taking into account} the 
    magnification effect in the theoretical template 
    ($C_{\ell}^{n_in_j}$), 
  \item $+g$: using galaxy counts alone {\it neglecting} the 
    magnification effect in the theoretical template 
    ($C_{\ell}^{g_ig_j}$), 
  \item $+n+\gamma$: combining galaxy counts, cosmic shear, and their 
    cross-correlations {\it taking into account} the effect of 
    magnification ($C_{\ell}^{n_in_j}$, $C_{\ell}^{\gamma_in_j}$, 
    $C_{\ell}^{\gamma_i\gamma_j}$), 
  \item $+g+\gamma$: combining galaxy counts, cosmic shear, and their 
    cross-correlations {\it neglecting} the effect of magnification, 
    ($C_{\ell}^{g_ig_j}$, $C_{\ell}^{\gamma_ig_j}$, 
    $C_{\ell}^{\gamma_i\gamma_j}$). 
\ei
Note that, in all cases, we add the primary CMB information (i.e., 
$C_{\ell}^{\Theta\Theta}$, $C_{\ell}^{\Theta E}$ and $C_{\ell}^{EE}$) in 
the Fisher matrix. The primary CMB power spectra are used to estimate 
the cosmological parameters up to $\ell=3000$. In Table \ref{const}, we 
show the forecast results of 1-$\sigma$ constraints on $\fnl$ for three 
representative surveys. Also, in Figs.\ref{HSC}-\ref{LSST}, 
two-dimensional contours of 1-$\sigma$ error on $b_z$-$\fnl$, 
$b_0$-$\fnl$ and $b_0$-$b_z$ planes are plotted in the cases of HSC, DES 
and LSST taking into account or neglecting the magnification effect. 

Let us focus on the results from the galaxy counts alone, i.e., $+n$ and 
$+g$ (see red solid and dashed lines in Figs.\ref{HSC}-\ref{LSST}). 
Naively, the statistical error on $\fnl$ is expected to be large if we 
properly take account of the magnification effect, because we need to 
specify the slope indices observationally and the number of parameters 
to be determined increases. As shown in Figs.\ref{HSC}-\ref{LSST}, 
however, the statistical error on $\fnl$ does not change so much in all 
cases, and the fractional change is around a sub-percent level 
(see Table.\ref{const}). These figures also show that, as mentioned in 
Sec.\ref{sec3}, the degeneracy between $\fnl$ and the galaxy bias 
($b_0$ and $b_z$) is weak (the correlation coefficient is $\sim 0.1$). 
On the other hand, the systematic bias on $\fnl$ arising from the 
incorrect treatment of the magnification effect is significant. As shown 
in Table.~\ref{const}, the systematic bias is apparently very large for 
DES. Taking the ratio of the systematic bias to the statistical error, 
however, the largest value is obtained from the LSST case amongst three 
surveys. That is, the systematic bias is rather serious for LSST than 
for HSC or DES. 

Next discuss the importance of cosmic shear information, i.e., 
$+n+\gamma$ and $+g+\gamma$ (see green solid and dashed lines in 
Figs.~\ref{HSC}-\ref{LSST}). Compared to the cases with galaxy counts 
alone, the addition of cosmic shear data would not only improve the 
statistical error, but also reduce the systematic bias on $\fnl$, 
irrespective of the treatment of the magnification effect. These results 
basically come from the non-vanishing shear-galaxy correlations, which 
also carry the information on $\fnl$ through the scale-dependent galaxy 
bias, as shown in Fig.~\ref{Cl}. Table.~\ref{const} shows that compared 
to the results with galaxy counts alone, the improvement of the 
constraint is by a factor of $\sim 2$ for LSST and of $\sim 1.3-1.5$ 
for HSC and DES. The size of systematic bias on $\fnl$ is now well 
within the 1-$\sigma$ statistical error for HST and DES, but it is still 
non-negligible for LSST.

\subsubsection{
Dependence on maximum multipole $\ell\rom{max}$, mean redshift $\zm$, 
number of redshift bin $N\rom{bin}$, and slope $s_i$
\label{Sec.4.2.2}
} 

\begin{figure}[t]
\bc
\includegraphics[width=85mm,clip]{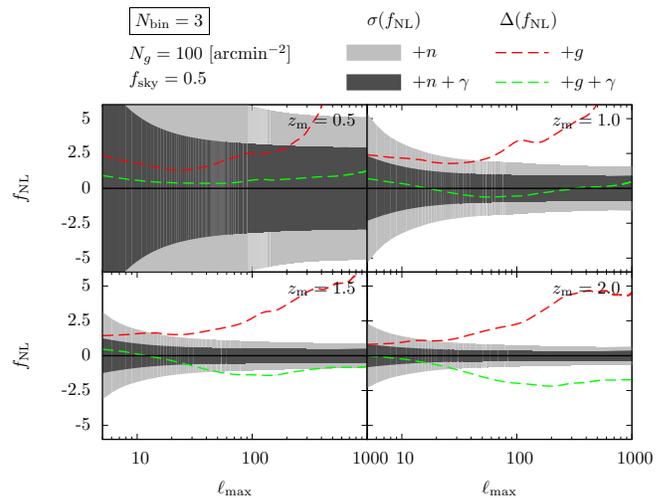} 
\ec 
\caption{
\label{LSSTn3}
The systematic bias on $\fnl$ as a function of $\ell\rom{max}$ for $+g$ 
(red) and $+g+\gamma$ (green). The shaded regions denote the 1-$\sigma$ 
constraint on $\fnl$ for $+n$ (thin) and $+n+\gamma$ (thick). Each panel 
show the case with $\zm=0.5$, $1.0$, $1.5$ and $2.0$. The galaxy 
subsamples are divided into three redshift bins ($N\rom{bin}=3$). We 
assume $\Ng=100$ arcmin$^{-2}$ and $\fsky=0.5$. 
}
\end{figure}

\begin{figure}[t]
\bc
\includegraphics[width=85mm,clip]{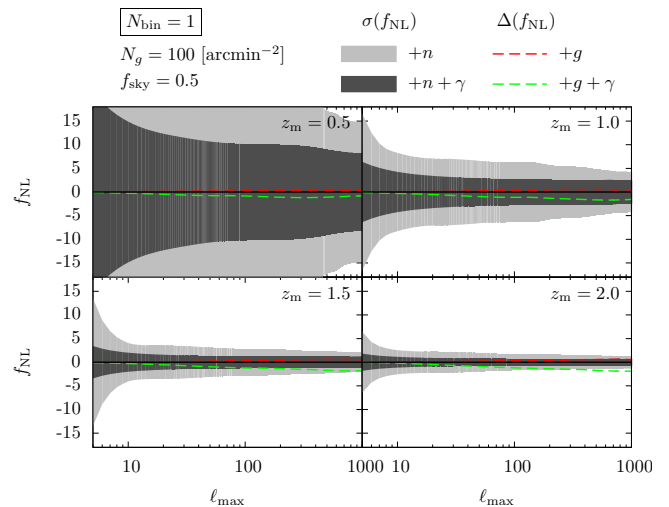}
\ec 
\caption{
Same as Fig.\ref{LSSTn3} but for $N\rom{bin}=1$. In this case, we choose 
the slope as $s_1=1.0$. Note that the range of $y$-axis is three times 
larger than that in Fig.\ref{LSSTn3}. 
\label{LSSTn1}
}
\end{figure}

To elucidate the results in Sec.\ref{Sec.4.2.1} in more details, we here 
consider the dependence of the systematic bias and statistical error on 
the various parameters, especially focusing on the LSST-like survey. 
Fig.~\ref{LSSTn3} shows the systematic bias $\Delta \fnl$ using the 
tomographic technique with $N_{\rm bin}=3$, plotted against the maximum 
multipole used in the parameter estimation, $\ell\rom{max}$. In each 
panel, the results obtained from a different survey depth are presented; 
$\zm=0.5$, $1.0$, $1.5$ and $2.0$ (in each case, redshift ranges of each 
redshift bin are summarized in Table.~\ref{zm}). To infer the 
significance of the systematic bias, we also plot the 1-$\sigma$ 
constraint on $\fnl$, $\sigma(\fnl)$, taking a proper account of the 
magnification effect (shaded region). Fig.~\ref{LSSTn1} also shows the 
same results as in Fig.~\ref{LSSTn3}, but, this time, we do not use the 
tomographic technique (i.e., $N\rom{bin}=1$), and the slope index is 
simply set to $s_1=1.0$. Note that the systematic bias does not depend 
on $\fsky$, while the 1-$\sigma$ constraint is proportional to 
$\sqrt{\fsky}$. Also, for a sufficiently large number density with 
$\Ng\gsim 10$ arcmin$^{-2}$, the results are almost insensitive to the 
choice of $\Ng$. Hence, the noise spectra are computed with fixed values 
of $\fsky=0.5$ and $\Ng=100$ arcmin$^{-2}$. 

Let us consider the results using the galaxy counts alone with 
$N\rom{bin}=3$ (see Fig.\ref{LSSTn3}). As increasing $\zm$, the 
statistical error on $\fnl$ (the thin shaded region) becomes rather 
improved, and the systematic bias (the red dashed lines) is somehow 
reduced if we choose a large maximum multipole. However, for a deep 
survey with $\zm\gsim1.5$ (bottom panels), the systematic bias is still 
non-negligible compared to the statistical error on $\fnl$. This is true 
even if we include the cosmic shear data (the green dashed lines).  

Next discuss the importance of tomographic technique. Comparison between 
Figs.~\ref{LSSTn1} and \ref{LSSTn3} implies that the tomographic 
technique is quite helpful to reduce the statistical error of $\fnl$. 
Typically, with $N_{\rm bin}=3$, the error would be reduced by a factor 
of $\sim 2-3$. This is partly because the amplitude of the large-scale 
galaxy clustering sensitively depends on the redshift in the presence of 
the scale-dependent galaxy bias (see Eq.~(\ref{Deltab}) and 
Fig.~\ref{Clz}). Presumably, this may also break the degeneracy between 
$\fnl$ and slope indices. Note, however, that the tomographic technique 
do not help to reduce the systematic bias, but rather it increases the 
size of bias. In this respect, a proper account of the magnification 
effect in the theoretical template is crucial in a LSST-like survey. 

\begin{figure}[t]
\bc
\includegraphics[width=85mm,clip]{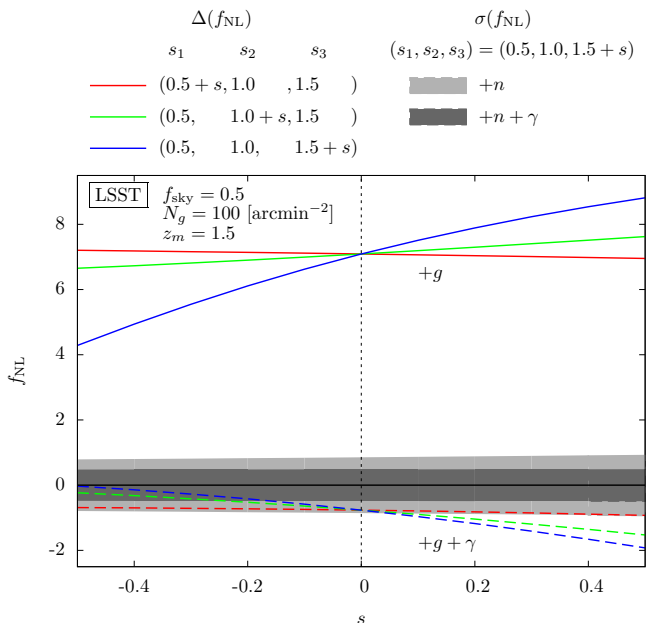}
\ec 
\caption{
Dependence of the systematic bias and 1-$\sigma$ constraint on the slope 
parameter. We vary only the one slope parameter ($s_1$, $s_2$ or $s_3$) 
and parametrized by a parameter $s$ as $s_1=0.5+s$, $s_2=1.0+s$, or 
$s_3=1.5+s$. The parameter $s$ is varied from $-0.5$ to $0.5$. The red, 
green and blue lines show the systematic bias with varying $s_1$, $s_2$ 
and $s_3$, respectively. The gray and black shaded regions represent the 
1-$\sigma$ constraint in the case of $+g$ and $+g+\gamma$, respectively, 
with varying the slope parameter in third redshift bin, $s_3$. We assume 
LSST to calculate the noise spectrum. 
\label{slope}
}
\end{figure}

So far, the fiducial values of the slope indices have been held fixed. 
In order to clarify the influence of the slope indices, keeping the 
number of redshift bins $N_{\rm bin}=3$, we allow to vary one of the 
three slope indices around the fiducial values, and estimate the 
systematic bias and statistical error on $\fnl$. Fig.~\ref{slope} shows 
the results in the case of the LSST survey. Here, the results are 
plotted against the shift of the slope index, $s$, defined by 
$s_i=s_{i,{\rm fid}}+s$. The lines with different color indicate the 
systematic bias of $\fnl$ obtained by varying the different slope index. 
Note that the statistical errors depicted as shaded regions are 
evaluated specifically in the case varying the slope $s_3$. 

Fig.~\ref{slope} indicates that the systematic bias of $\fnl$ is 
insensitive to the variation of slope index in lower redshift bin, but 
rather sensitive to it at higher redshift bin. This result is quite 
reasonable because the magnification effect at high-$z$ bin is 
significant compared to low-$z$ bin (see Fig~\ref{Clz}). On the other 
hand, the statistical error on $\fnl$ is insensitive to the variation 
of slope index, as it is expected from Sec.~\ref{sec3.2}. In this 
respect, depending on the value of slope indices at high redshifts, the 
magnification effect on constraining primordial non-Gaussianity may 
become even more serious, and again, should be properly taken into 
account in the theoretical template.

\section{
\label{sec5}
Summary 
}

In this paper, we studied the impact of magnification effect on the 
detection of $\fnl$ from photometric survey. As representative upcoming 
photometric surveys, we considered HSC for deep, and DES for wide, and 
LSST for an idealistically deep and wide survey. From the Fisher matrix 
analysis, we showed that, an incorrect treatment of the magnification 
effect on the theoretical template of angular power spectra leads to the 
systematic bias in the best-fit value of $\fnl$. Especially, using galaxy 
counts alone, the size of systematic bias is significant for all three 
surveys (HSC, DES and LSST), and true values of $\fnl$ would typically 
go outside the $3$-$\sigma$ error of the biased confidence region. 
However, we found that additional information from the cosmic shear 
observations helps not only to improve the constraint, but also to 
reduce the systematic bias on $\fnl$. As a result, the systematic bias 
can become negligible for HSC and DES surveys, compared to their 
expected errors on $\fnl$. A proper account of the magnification effect 
does not increase the statistical error on $\fnl$ ($\lsim1$\% ). 
Nevertheless, for LSST, a relative significance of the systematic bias 
still remains and the magnification effect should be correctly taken 
into account in the theoretical treatment.  

We further explored the various cases by changing parameters 
characterizing the survey properties, and showed that the tomographic 
technique using photometric redshift information leads to a significant 
improvement on the statistical error on $\fnl$, but it doe not help to 
reduce the systematic bias. In any case, high-$z$ observations are 
indispensable for tightly constraining primordial non-Gaussianity, but 
the influence of the magnification effect would be inevitable. This is 
particularly true for deep imaging surveys like LSST ($\zm\gsim1.5$). 
A proper account of the magnification effect in the theoretical 
template is thus quite essential for an unbiased estimate of primordial 
non-Gaussianity.


\begin{acknowledgments}
We would like to thank Emiliano Sefusatti and Fabian Schmidt for 
useful comments and discussions about the size bias. We are also 
grateful to Eiichiro Komatsu and Toshifumi Futamase for helpful 
discussions. TO and AT are supported in part by a 
Grants-in-Aid for Scientific Research from the Japan Society for the 
Promotion of Science (JSPS) (No. 22-2879 for TO and No. 21740168 for AT). 
This work was supported in part by Grant-in-Aid for Scientific Research 
on Priority Areas No. 467 "Probing the Dark Energy through an Extremely 
Wide and Deep Survey with Subaru Telescope", the GCOE Program "Weaving 
Science Web beyond Particle-matter Hierarchy" at Tohoku University, and 
JSPS Core-to-Core Program "International Research Network for Dark 
Energy".
\end{acknowledgments}

\appendix


\end{document}